\documentclass[a4paper,12pt,twocolumn]{article}
\usepackage{cite}
\usepackage[dvips,hiresbb]{graphicx}
\usepackage{here}

\title{\textbf{Special polarization characteristic features of
a three-dimensional terahertz photonic crystal
with a silicon inverse diamond structure}}
\author{\Large{Chikara Sakurai} \thanks{Email: c-sakurai@river-ele.co.jp (first), sakuraikazan@gmail.com (second)} \\ \textit{River Electric Corporation, 2-1-11 Fujimigaoka, Nirasaki, Yamanashi, Japan}} 
\date{\today}
\begin{document}
\onecolumn{
\maketitle
\begin{abstract}
\normalsize {The band structure of an Si inverse diamond structure whose lattice point shape was vacant regular octahedrons, was calculated using plane wave expansion method.  A complete photonic band gap was theoretically confirmed at around 0.4 THz. 
It is said that three-dimensional photonic crystals have no polarization anisotropy in photonic band gap (stop gap, stop band) of high symmetry points in normal incidence. However, it was experimentally confirmed that the polarization orientation of a reflected wave was different from that of an incident wave, [I$(X,Y)$], where $(X,Y)$ is the coordinate system fixed in the photonic crystal.
It was studied on a plane (001) at around X point's photonic band gap (0.36 $-$ 0.44 THz) for incident wave direction [001]  by rotating a sample in the plane (001), relatively.
The polarization orientation of the reflected wave was parallel to that of the incident wave when that of the incident wave was I(1, 1) or I(1, $-$1). In contrast, the former was perpendicular to the latter when that of the incident wave was I(1, 0) or I(0, $-$1) at around 0.38 THz.
As far as the photonic crystal in this work is concerned, method of resolution and synthesis of the incident polarization vector is not able to apply to the analyses of rotation of the measured reflected spectra in appearance.} 
\end{abstract}
\twocolumn{
\section*{Introduction}
\hspace*{5mm}Recently, THz (terahertz) technologies, the range of which is located halfway between microwaves and infrared lights, have been proceeding in various industrial fields such as the body checks for securities, cameras~\cite{oda}, chemical identifications~\cite{hos}, non-destructive examinations~\cite{dob, kaw, ari}, and microscopes~\cite{sal1} etc..\\
\hspace*{5mm}The three-dimensional (3D) photonic crystals (PC) have the regular periodicity of dielectric materials~\cite{yab,ozb,nod,kwa}. The 3D-PC can control waves by forming point defects and line defects on the surface and in the interior and they have possibilities of further contribution to photonic devices and new characteristic features. The 3D-PC may also present a new frontier for them.\\
\hspace*{5mm}In this work, the band structure of an Si inverse diamond structure was calculated using plane wave expansion method.  The lattice point shape of this structure was vacant regular octahedrons and it was theoretically confirmed that the complete photonic band gap (CPB) existed at around 0.4 THz.\\
\hspace*{5mm}The polarization anisotropy, which was the polarization orientation (electric-field direction) difference between a reflected wave and incident one, was studied on the surface (001) at around BGX that is X point's photonic band gap. The direction of the incident wave is Z-direction, [001].  The polarization anisotropy of the reflected wave was studied with four kinds of polarization orientation of the incident wave, [I($X,Y$)] on the surface (X-Y plane). They were I(1, 1), I(1, 0), I(1, $-$1), and I(0, $-$1).\\
\hspace*{5mm}The polarization orientation of measured reflected spectra was rotated by 90 degrees at a frequency within BGX when that of the incident wave was I(1, 0) or I(0, $-$1).
\section*{Photonic Band Structure}
\hspace*{5mm}The lattice of the diamond structure is shown in fig.~\ref{fig:uclps}(a). The sphere is the lattice point and its shape is the regular octahedrons as shown in fig.~\ref{fig:uclps}(b).
\begin{figure}[h]
\centering
\includegraphics[width=70mm,]{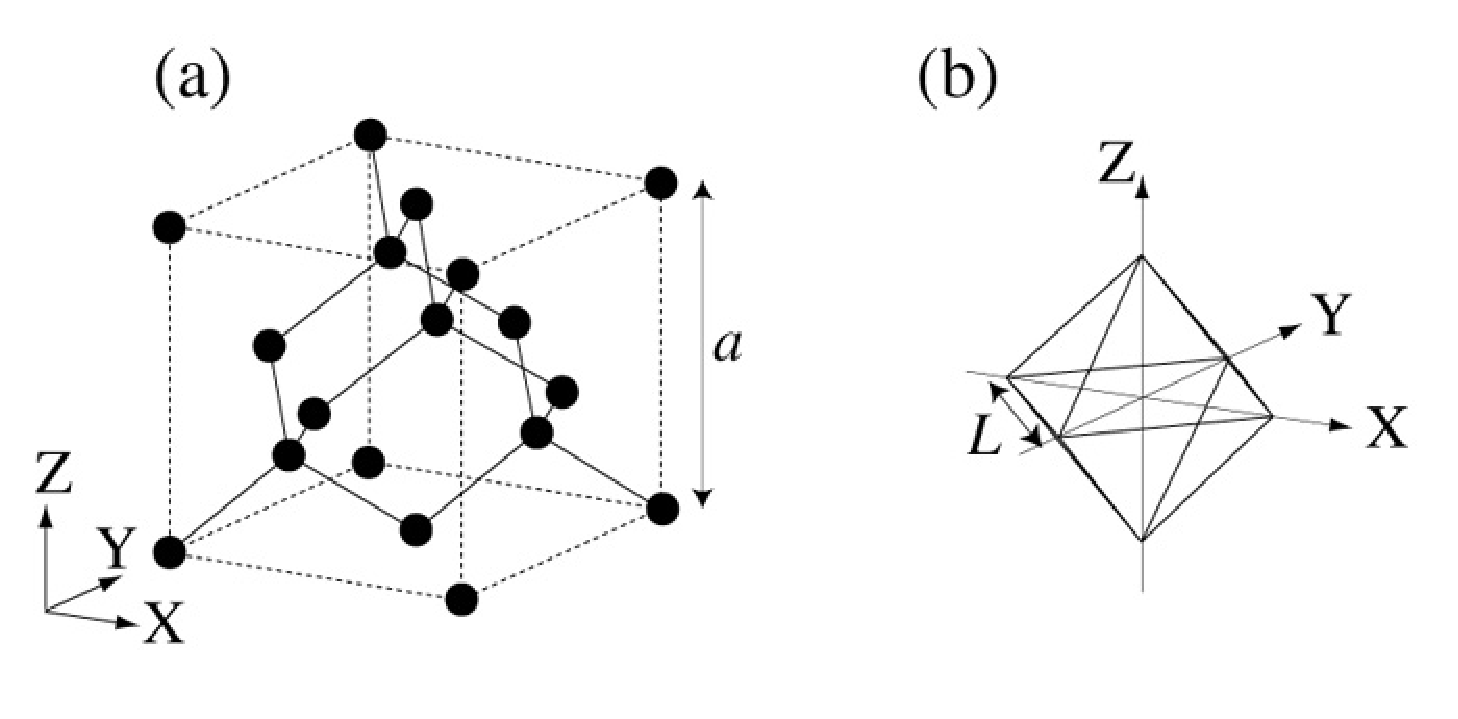}
\caption{\footnotesize (a) Lattice of the diamond structure. (b) Shape of the lattice point is the regular octahedrons. It is vacant ($\varepsilon_\mathrm{a}$ = 1.00). The surrounding material is Si  ($\varepsilon_\mathrm{b}$ = 11.9). }
\label{fig:uclps}
\end{figure}
It is vacant and the dielectric constant, $\varepsilon_\mathrm{a}$ is 1.00 (atmosphere). 
The surrounding material is pure Si (resistivity, $\rho >10^4\ \Omega$ cm) and the dielectric constant, $\varepsilon_\mathrm{b}$ is set as 11.9.
The lattice constant, a  is 300 $\mu$m and the length of the regular octahedrons side, L is set as 150 $\mu$m in this theoretical and experimental works.\\
\hspace*{5mm}Fig.~\ref{fig:bsbz}(a) shows the calculated photonic band structure using plane wave expansion method, and CPB exists at around 0.4 THz.
The first Brillouin zone is shown in fig.~\ref{fig:bsbz}(b).\\
\hspace*{5mm}The direction of the incident wave is Z-direction, [001]
\footnote{[001], (001), \{100\}, I($X,Y$) and so on are defined on the X-Y-Z coordinate system in fig.~\ref{fig:uclps}(a).}
in the real space and it corresponds to $\Gamma$-X direction in the wave number space (K-space). BGX exists between 0.36 and 0.44 THz.
\\
\begin{figure}[t]
\centering
\includegraphics[width=70mm,]{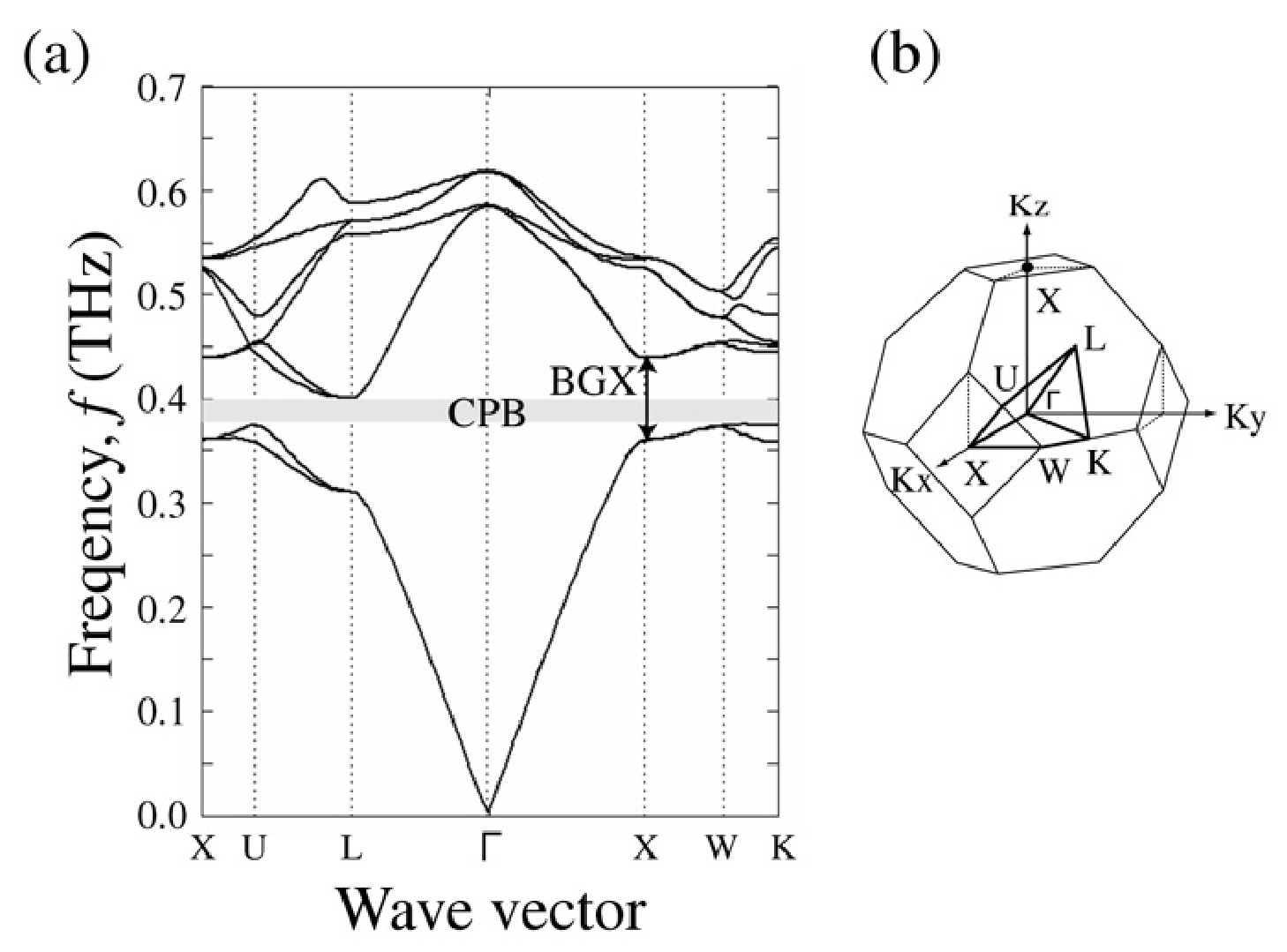}
\caption{\footnotesize (a) Calculated photonic band structure has CPB (gray zone) at around 0.4 THz. BGX exists between 0.36 and 0.44 THz. (b) First Brillouin zone and the reduced zone (heavy line) with high symmetry points.}
\label{fig:bsbz}
\end{figure}
\section*{Experimental System}
\hspace*{5mm}The main periodic square patterns (side $L = 150$ $\mu$m) were etched on both surfaces (001) of an Si plate whose height was 75 $\mu$m and area was $10\times10$ mm$^2$ and  and they were periodically arranged along the direction [001]
\footnote{A stereograph of the sample and 3D figure of the unit cell are illustrated in detail in arXiv:1801.06741.}. \\
\hspace*{5mm}The etching angle of Si \{100\} surface is 54.7 degrees and the etched vacant shape forms the regular octahedrons between four layers. The height of four layers is 300
$\mu$m, and it corresponds to lattice constant, a.
The total number of the layered Si chips was 48 layers whose height was 12a.\\
\hspace*{5mm}The schematic diagram of the measurement system with THz-TDS (time-domain spectroscopy) equipment owned by Nippo Precision co., ltd. is shown in fig.~\ref{fig:ms}.
\begin{figure}
\centering
\includegraphics[width=70mm,]{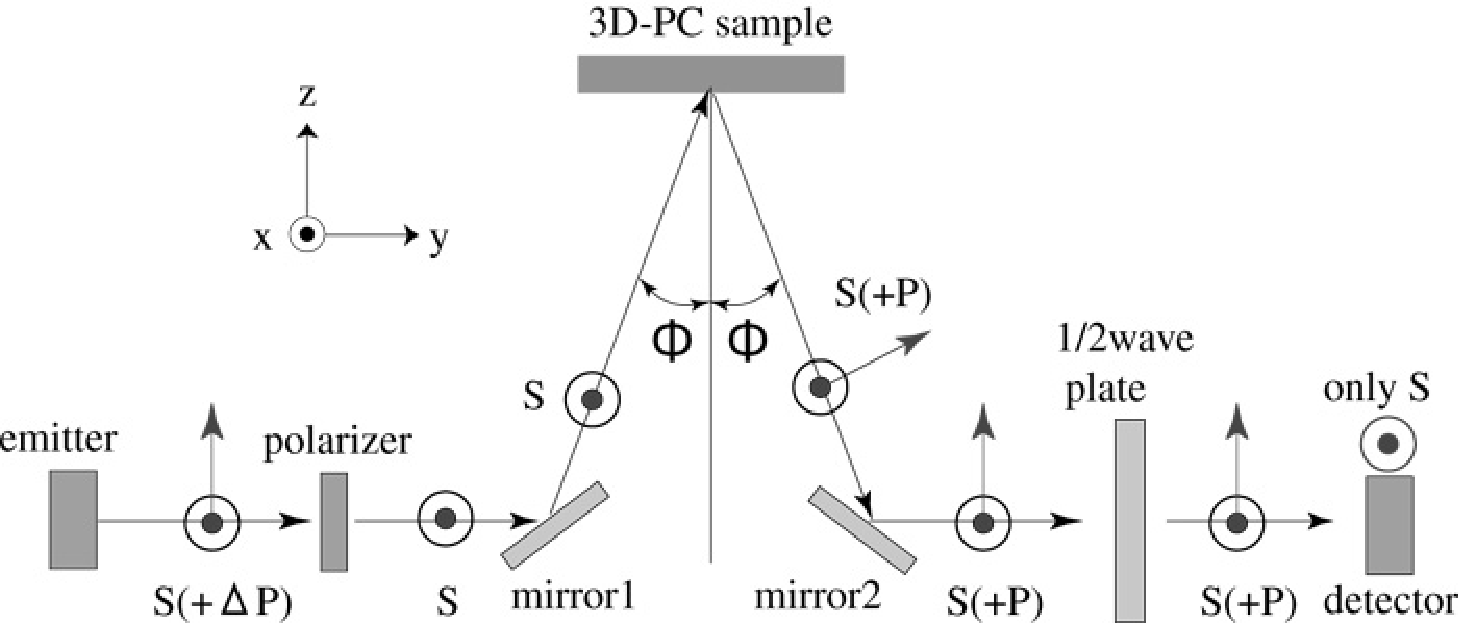}
\caption{\footnotesize Schematic diagram of the measurement system with the THz-TDS. The 3D-PC sample is set as the layered direction $\|$ z. The incident angle, $\phi$  is 7 degrees. The polarization orientation of the emitter is S-p $\|$ x and the detector detects only S-p. The designed 1/2 wave plate converts S-p into P-p and P-p into S-p at around 0.4 THz. S(+P) means S-p, P-p or the mixing of S-p and P-p.}
\label{fig:ms}
\end{figure}
The polarization orientation of the emitter, which is fixed, is S polarization (S-p) and it is parallel to x-axis.
The mirror 1 and mirror 2 are coated with Au.
S-p launched by the emitter is reflected by the mirror 1 and the polarization orientation is also S-p \footnote{The expression of the inversion, phase shifting by $\pi$, of S-p is abbreviated in fig.~\ref{fig:ms}.}.\\
\hspace*{5mm}The 3D-PC sample is so horizontally set that the layered direction is parallel to z-axis, which corresponds to Z-axis in fig.~\ref{fig:uclps}.
The incident angle is 7 degrees, which is the normal incidence approximately.\\
\hspace*{5mm}Another polarization perpendicular to\\ S-p is P polarization (P-p) that is included in the incidence plane.
According to Fermat's principle, the reflection angle is equal to the incident one. When not only S-p but P-p is included in the reflected wave of the sample, S-p and P-p are naturally included in the reflected wave by the mirror 2.\\
\hspace*{5mm}Meanwhile, the detector detects only \\S-p, which is also fixed. Therefore, a 1/2 wave plate was used for measurements of \\
P-p component part included in the reflected wave of the sample.\\
\hspace*{5mm}The material of the 1/2 wave plate is quartz, SiO$_2$. Xc and Zc (crystal axes) are included in the plane of the plate and Yc is parallel to the direction of the thickness. The ordinary and extraordinary refractive indices are $n_\mathrm{{o}}=n_\mathrm{{Xc}}=n_\mathrm{{Yc}}=2.108$ and $n_\mathrm{e}=n_\mathrm{{Zc}}= 2.156$ at 1 THz, respectively~\cite{mas}.\\
\begin{figure}[t]
\centering
\includegraphics[width=40mm,]{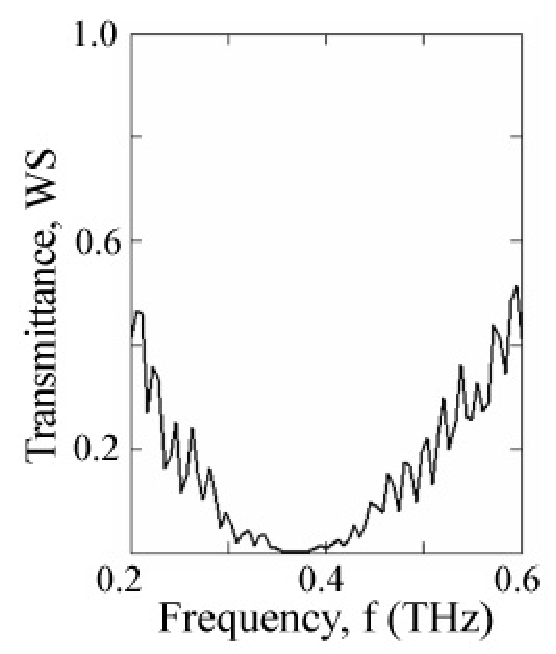}
\caption{\footnotesize Transmittance, $WS(f)$ of the designed 1/2 wave plate at around 0.4 THz. For example, S-p incident wave is nearly converted into P-p at 0.37 THz though the 1/2 wave plate.}
\label{fig:4ab}
\end{figure}
\hspace*{5mm}Yc is parallel to y-axis. The designed 1/2 wave plate, whose thickness is 8.45 $mm$, converts S-p into P-p and P-p into S-p at around 0.4 THz when one of two bisector of Xc and Zc is parallel to x-axis.
Meanwhile, when one of Xc or Zc  is parallel to x-axis, the orientation of the incident polarization, S-p or P-p does not change though the plate. \\
\hspace*{5mm}$avXZ(f)$ is defined as the average of transmitted spectra of S-p $\|$ Xc and S-p $\|$ Zc. Variable, $f$ is terahertz (THz) frequency. It was used for the normalization of transparent characteristics of the 1/2 wave plate. \\
\hspace*{5mm}Fig.~\ref{fig:4ab} shows the transmittance, $WS(f)$\footnote{The concavity and convexity in fig.~\ref{fig:4ab} appear from the reflection of back surface of the 1/2 wave plate.} 
which is the spectrum normalized by $avXZ$ though the 1/2 wave plate at around BGX.
$1-WS(f)$ is the ratio of P-p conversion into S-p, inversely. 
In measurement, the polarization orientation of the incident wave was S-p and an Au plate was set instead of the 3D-PC sample.\\
\hspace*{5mm}$WS(f)$ was used as a reference of the reflected spectra 
when the spectra were measured by using the 1/2 wave plate .\\
\section*{Experimental Results and Discussions}
\hspace*{5mm} The sample was rotated in the x-y plane in fig.~\ref{fig:ms} as a substitute for the rotation of S-p of the incident wave for measurements of polarization anisotropy
of the reflected wave. As stated above, z-axis corresponds to Z-axis in fig.~\ref{fig:uclps}.\\
\hspace*{5mm}The polarization orientation of the incident wave, I$(X,Y)$ is defined as vector ${(X,Y,0)}$.
For example, it is called I(1, 1) when S-p of the incident wave is parallel to ${(1,1,0)}$. The polarization spectra of the reflected wave were measured for four kinds of orientation that
were [I(1,~1), I(1,~0), I(1,~$-$1), I(0,~$-$1)]
\footnote{For example, I(1, 0) and I($-$1,0) are identical since two directions are indistinguishable in measurement. }.\\
\begin{figure}[t]
\centering
\includegraphics[width=71mm,]{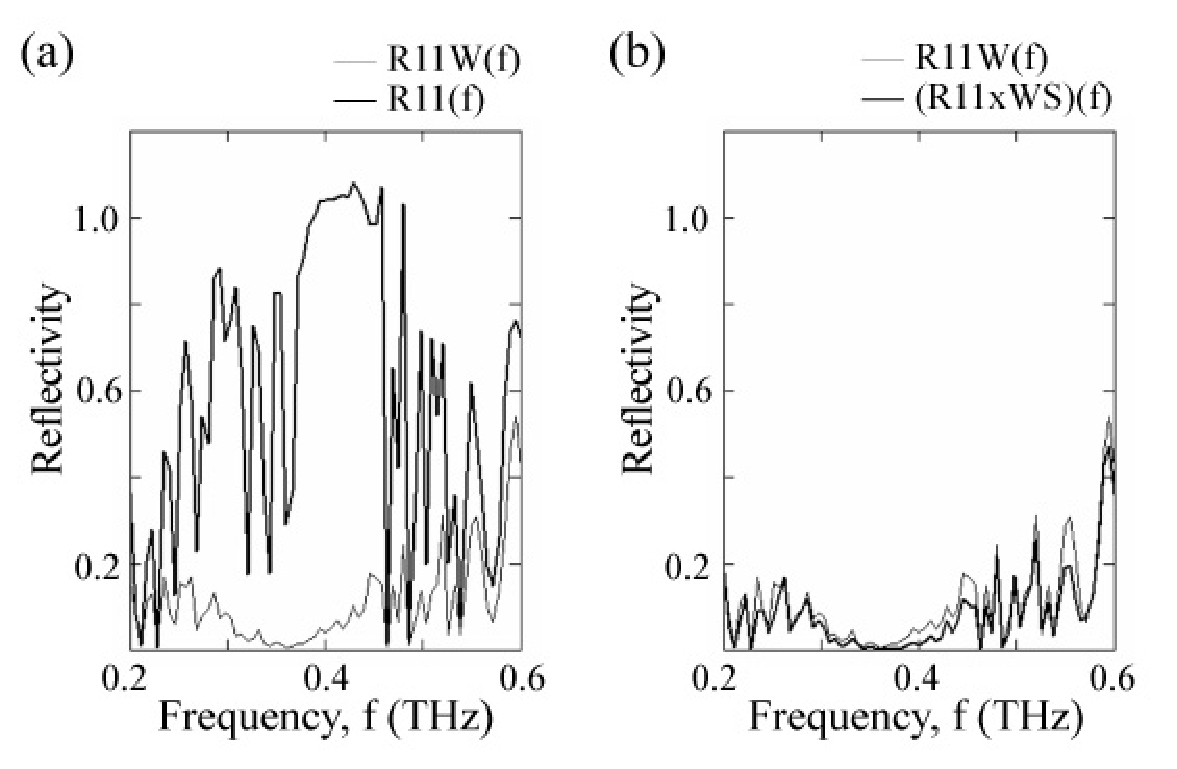}
\caption{\footnotesize Reflected spectra measured with I(1, 1) at around BGX.  
(a) heavy line: $R11(f)$ with no 1/2 wave plate. thin line: $R11W(f)$ with one.
(b) $R11(f)\times WS(f)$ is nearly equal to $R11W(f)$. It means that the polarization orientation of the reflected spectra is nearly equal to S-p orientation of the incident wave.}
\label{fig:5ab}
\end{figure}
\hspace*{5mm}Fig.~\ref{fig:5ab} shows the reflected spectra measured with I(1, 1) at around BGX. In fig.~\ref{fig:5ab}(a), $R11(f)$ is the reflectivity  with no 1/2 wave plate and it is normalized by Au reflected spectra. $R11W(f)$ is the reflectivity  though the 1/2 wave plate and it is normalized by $avXZ$. The reflected spectra with I(1, $-$1) also have similar characteristics.
 In fig.~\ref{fig:5ab}(b),
$R11(f)\times WS(f)$ is nearly equal to $R11W(f)$.
Therefore, this experimental results indicate that $R11(f)$ consists of only S-p.\\
\begin{figure}[t]
\centering
\includegraphics[width=76mm, trim=0 0 0 4]{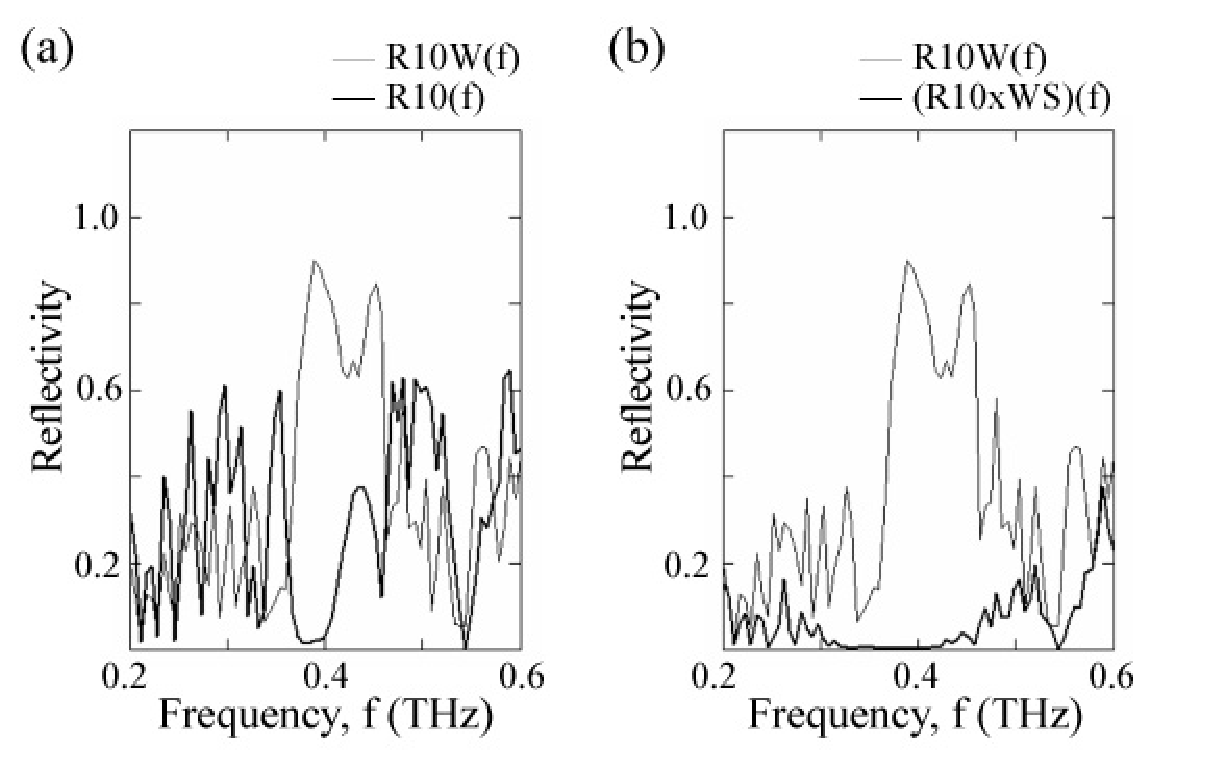}
\caption{\footnotesize Reflection spectra measured with I(1, 0) at around 0.4 THz (BGX). 
(a) heavy line: $R10(f)$ with no 1/2 wave plate. thin line: $R10W(f)$ with one.  
(b) $R10(f)\times WS(f)$ is entirely different from $R10W(f)$.}
\label{fig:6ab}
\end{figure}
\begin{figure}[h]
\centering
\includegraphics[width=40mm,]{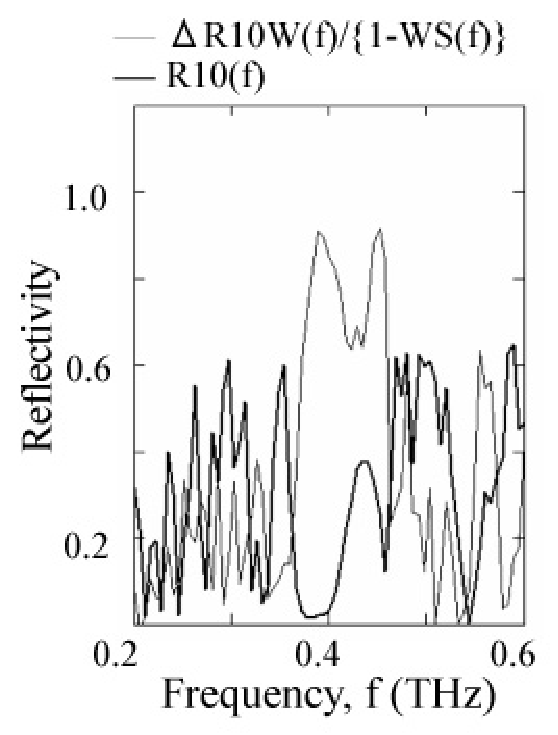}
\caption{\footnotesize $\Delta R10W(f)/[1-WS(f)]$ and $R10(f)$ corresponds to P-p and S-p reflected spectra, respectively with no 1/2 wave plate when the incident wave S-p is I(1, 0).  Especially at around 0.38 THz within BGX, S-p incident wave is almost entirely converted into P-p reflected wave.}
\label{fig:7ab}
\end{figure}
\hspace*{5mm}Fig.~\ref{fig:6ab} shows the reflected spectra measured with I(1, 0) at around BGX. In fig.~\ref{fig:6ab}(a), $R10(f)$ is the reflectivity  with no 1/2 wave plate and it is normalized by Au reflected spectra. $R10W(f)$ is the reflectivity  though the 1/2 wave plate and it is normalized by $avXZ$.  The reflected spectra of I(0, $-$1)  also have similar characteristics.
In fig.~\ref{fig:6ab}(b), $R10(f)\times WS(f)$ is not entirely  equal to $R10W(f)$ at around BGX.  It means that the reflected spectra includes P-p besides S-p.\\
\hspace*{5mm}In fig.~\ref{fig:7ab}, $\Delta R10W(f)$ that is the difference  between $R10W(f)$ and $R10(f)\times WS(f)$, corresponds to the contribution from P-p though the 1/2 wave plate. Accordingly, the normalized $\Delta R10W(f)/[1-WS(f)]$ corresponds to P-p reflected spectrum with no 1/2 wave plate. Especially at around 0.38 THz, S-p incident wave is almost entirely converted into P-p reflected wave.\\
\hspace*{5mm}Though the two directions in XY-plane,
(1, 1) and (1, $-$1) are identical in appearance in the diamond structure,  the polarization orientation of the reflected wave was rotated by 90 degrees to the polarization orientation
of the incident wave, I(1, 0). These results suggest that method of resolution and synthesis of the incident polarization vector is not able to apply in appearance\footnote{In arXiv:1811.02990, the polarization rotation of the reflected wave is
explained by using phase analyses with a FEM (finite element method).
The fundamental causation of the phenomenon, however, remains unsolved.}.\\
\section*{Conclusions}
\hspace*{5mm}The Si inverse diamond structure whose lattice point shape was vacant regular octahedrons was calculated with plane wave expansion method, and it  had the complete photonic band gap at around 0.4 THz. 
The polarization anisotropy, which means that the polarization orientation of a reflected wave is different from that of an incident wave, was experimentally studied at around BGX (0.36 $-$ 0.44 THz) on the plane (001). The incident wave direction was [001] which corresponds to $\Gamma$-X direction and 
the kind of polarization orientation of the incident wave was [I(1,~1), I(1,~0), I(1,~$-$1), I(0,~$-$1)] in the X-Y plane}. The polarization orientation of the reflected wave was parallel to that of the incident wave at around BGX that is X point's photonic band gap for  I(1,~1) and I(1,~$-$1). In contrast, the polarization orientation of the reflected wave was entirely different from that of the incident wave for I(1, 0) and I(0, $-$1). Especially, the former was perpendicular to the latter at around 0.38 THz within BGX.
As far as the three-dimensional photonic crystal in this work is concerned, the method of resolution and synthesis of the incident polarization vector is not apparently able to apply to the analyses of rotation of the measured reflected spectra.
\\
\section*{Acknowledgment}
\hspace*{5mm}The author would like to thank Hidekazu Onishi, who fabricated the designed Si inverse diamond structure and Ph.D. Takeshi Sawada, who is
a research scientist, Terahertz Project, Second Design Department, Nippo Precision co., ltd.
\\

}
\end{document}